\documentclass[preprint,12pt,authoryear]{elsarticle}

\usepackage{amssymb}
\journal{Journal of Quantitative Spectroscopy \& Radiative Transfer}

\begin{document}

\begin{frontmatter}



\title{Testing sky brightness models against radial dependency: a
    dense two dimensional survey around the city of Madrid, Spain}


\author[ucm]{J. Zamorano\corref{cor}}
\ead{jzamorano@fis.ucm.es}

\author[ucm]{A. S\'anchez de Miguel\corref{cor}}
\author[ucm]{F. Oca\~na}
\author[ucm]{B. Pila-D\'iez}
\author[ucm]{J. G\'omez Casta{\~n}o}
\author[ucm]{S. Pascual}
\author[ucm]{C. Tapia}
\author[ucm]{J. Gallego}
\author[ucm]{A. Fern\'andez}
\author[ucm]{M. Nievas}

\address[ucm]{Dept. Astrof\'isica y CC. de la Atm\'osfera, Universidad Complutense de Madrid, \\Ciudad Universitaria, 28040 Madrid, Spain}

\cortext[cor]{These authors contributed equally to this work}

\begin{abstract}
We present a study of the night sky brightness around the extended
metropolitan area of Madrid using Sky Quality Meter (SQM)
photometers. The map is the first to cover the spatial distribution of the sky brightness in the
center of the Iberian peninsula. These surveys are neccessary to test the
light pollution models that predict night sky brightness as a function
of the location and brightness of the sources of light pollution and
the scattering of light in the atmosphere. We describe the data-retrieval methodology, which
includes an automated procedure to measure from a moving vehicle in
order to speed up the data collection, providing a denser and wider
survey than previous works with similar time frames. We compare the night sky
brightness map to the nocturnal radiance measured from space by the
DMSP satellite. We find that i) a single
source model is not enough to explain the radial evolution of the
night sky brightness, despite the predominance of Madrid in size and
population, and ii) that the orography of the region should be taken
into account when deriving geo-specific models from general
first-principles models. We show the tight relationship between these two
luminance measures. This finding sets up an alternative roadmap to
extended studies over the globe that will not require the local
deployment of photometers or trained personnel.
\end{abstract}


\end{frontmatter}


\section{Introduction}\label{intro}
Light pollution (the introduction by humans, directly or indirectly,
of artificial light into the environment) is a major issue worldwide,
especially in urban areas. It increases the sky glow and prevents us
from observing a dark starry sky. This is why astronomers are among
the worst affected by urban sky glow \citep{cinzano2000artificial} and
they were probably the first to fight against light pollution. One of
the key parameters to select a site to build an observatory is the
night sky brightness because some astronomical research could not be
performed with the required quality if the sky is not dark
enough. Thus, it is not a surprise to find that the astronomical
observatories are located in remote areas far from light pollution
sources.

There are citizen campaigns in defense of the values associated with
the night sky and the general right of the citizens to observe the
stars. 'Starlight, A Common Heritage' promoted by the IAU and the
UNESCO, said: 'An unpolluted night sky that allows the enjoyment and
contemplation of the firmament should be considered an inalienable
right of humankind equivalent to all other environmental, social, and
cultural rights, due to its impact on the development of all peoples
and on the conservation of biodiversity'
\citep{starlight2007declaration}.

The main data input for artificial lighting registered from space has
been the images obtained with sensors onboard the US Air Force Defense
Meteorological Satellite Program ({\it DMSP}) Operational Linescan
System ({\it OLS}) developed to map human settlements. Using the
digital archive for {\it DMSP/OLS} data (available at the National
Geophysical Data Center), that contains annual cloud-free composites
of nighttime lights, it is possible to obtain a spatial model of
population density \citep{sutton1997comparison,sutton2003scale}, and
economic activity \citep{sutton2007estimation}. This archive is a very
useful source of data to study the evolution of light emission. An
expansion of lighting surrounding urban centres and areas of new
lighting are found \citep{elvidge2007change} although a reduction of
the sky glow that surrounds some big cities has been also found. This
good news may be the result of changes in lighting type or the
installation of lighting fixtures to direct light downwards. It is
worth mentioning that an application of these data consists in the
modeling of artificial night sky brightness and its effect on the
visibility of astronomical features
\citep{falchi1998maps,cinzano2000artificial,cinzano2000growth,
  cinzano2001naked,cinzano2001first,cinzano2004night}.

Urban illumination comes mainly from public lighting of the streets
and buildings. Detecting light pollution is straightforward by visual
inspection of the images which speak by themselves and are very useful
to draw public attention of the problem. The intensity of the pixels
is related to the amount of light being sent to space and scattered by
the atmosphere. The bright spots of light reveal an excess or bad use
of lighting. The extension and intensity of this emission put in
evidence that light pollution is, besides a concern for astronomers, a
global problem that is damaging our
environment\citep{longcore2004ecological}.

In this paper we address the study of light pollution and its effect
on the night sky brightness and on the visibility of the stars
\citep[see for instance][]{cinzano2004night} using night sky
brightness data collected from photometers on top of moving vehicles
around Madrid. We have used Sky Quality Meter (SQM) devices which are
pocket size photometers designed to provide measures of the luminance
or surface brightness of the sky (night sky brightness, NSB for short)
in astronomical units of magnitude per square arcsecond
($mag\;arcsec^{-2}$) \citep{cinzano2005night}. The resulting NSB map
(5389 $km^2$) is compared with calibrated images of radiance as
measured from space. Our test has been carried out in a region around
the big city of Madrid ($\sim5.4$ millions of inhabitants inside an
area of 27 km radius) for a total population in the region of around
6.5 millions of inhabitants.

A similar study was performed by \cite{biggs2012measuring} around the
city of Perth (Western Australia) with 1.27 millions of
inhabitants. Perth is a very isolated city and the NSB measurements
were not affected by light from nearby large cities. On the other hand
the measurements were made using hand held photometers with a total of
310 useful data points. The observations for the Hong Kong light
pollution map \citep{pun2012night} were performed by 170 volunteers
who acquired 1957 night sky measurements in 199 locations. The map
covers an area of 1100 km$^2$. To speed up the data acquisition,
\cite{espey2014initial} used a method based on measurements taken from
a vehicle and GPS information. They surveyed rural areas close to
Dublin (Ireland) with 1.27 millions of inhabitants.

One of the main motivations of this study has been to quantify the
increase of the night sky brightness as a consequence of the
artificial lighting at present around a big city. The results thus
become the reference values to compare with similar studies to be
carried out in the future that focus on the evolution of the light
pollution in the region around Madrid. The calibration and procedures
could be extended to the study of wider areas.

\section[]{Brightness of the Night Sky}
The history of the artificial sky glow measurements has been recently
reviewed by \cite{scikezor2013new}. To summarize, there are two
methods used by people joining citizen-science projects. On the one
hand, they report the number of stars that they could see after visual
observations of selected sky fields. These data inform about the
limiting magnitude of the sky, which is closely related to the sky
brightness. Although this method relies on subjective measures, it has
been shown to yield scientific information and its precision increases
when the number of observations increases \citep{kyba2015worldwide}.
On the other hand, people are using Sky Quality Meters (SQM), which
are hand held photometers that provide the night sky brightness in
astronomical units of $mag\;arcsec^{-2}$ in a band that includes the
{\it B} and {\it V} Johnson astronomical bands. The SQM is a portable
and easy to use device that could measure even in the darkest
skies. The report by \cite{cinzano2005night} showed that it could be
used to scientific purposes with an accuracy of 10\% between different
devices.

Professional scientists have been using images taken at night from
satellites to detect and to measure the sources of light pollution and
to estimate the sky brightness using models which take into account
the scattering of light by molecules and aerosols of the
atmosphere. This method provides a global estimation of the night sky
brightness
\citep[i.e.][]{falchi1998maps,kocifaj2007light,aube2012using}.

Astronomical spectroscopic observations can be used to determine not
only the amount of light pollution but also the type of luminaries
employed in public outdoor lighting. The bright emission lines of low-
and high-pressure sodium lamps, and also metal halide and compact
fluorescent lamps are the main features of the spectra of the light
polluted skies \citep{sanchezdemiguel2015variacion}.

Image observations to measure the magnitude of astronomical objects
are performed using standard astronomical photometry methods. The
images are taken through a filter and registered by a CCD
detector. Its quantum efficiency or spectral response, in conjunction
to the filter transmission, determines the photometric band. The
contribution of the background sky can be estimated in the nearby
field free of objects registered in the image and it should be
subtracted to obtain the net flux of the object. When processing
astronomical photometry observations the value of the background sky
is usually not stored after the analysis of the images. Since the
astronomers consider the sky background a subproduct, their values are
not published in the scientific papers. To get information on sky
background one has to browse archival data, to get the images, and
measure each one \citep{patat2003ubvri,patat2008dancing}.

It should be noted the difference between sky brightness and sky
background brightness. Measurements with the SQM photometer, for
instance, include the contribution of the stars in the field of view
of the device ($\sim$20 degrees FWHM). With CCD photometry a small
area of the sky free of stars is measured.

\section[]{Sky Brightness Data}
\subsection[]{Sky Quality Meters}
The Sky Quality Meter (SQM) is a pocket size photometer designed to
provide measurements of the luminance or surface brightness of the sky
(night sky brightness, NSB for short) in astronomical units of
magnitude per square arcsecond ($mag\;arcsec^{-2}$)
\citep{cinzano2005night}. This is a logarithmic unit with a scale
where an increase of 5 magnitudes is equivalent to a decrease of 100
in luminance. The surface brightness in $mag\;arcsec^{-2}$ is
equivalent to $-10 - 2.5 \times log_{10}$ (surface brightness in
photons $s^{-1} m^{-2} Hz^{-1} arcsec^{-2}$).

The photosensitive element is a silicon photodiode (ams-TAOS TSL237S) light-to-frequency converter covered by a HOYA CM-500 filter with final response covering the Johnson {\it B} and {\it V} bands used in astronomical photometry (wavelength range 320 to 720 nm). Although its appearance is simple and it is very user friendly (you should aim the photometer, push the button, and read the data on the display), it has been shown that it is accurate enough to perform scientific research \citep{cinzano2005night}. The SQMs have a quoted systematic uncertainty of $10\%$ ($0.10 \;mag\;arcsec^{-2}$). Its use has become very popular among researchers and amateur astronomers along with interested members of associations that fight against light pollution.

In this study we have used three models of photometer. The SQM-L, the simplest one, is intended for taking data in the field, aiming the photometer to the zenith for instance, using a photographic tripod. For mapping an area you should move to the selected points of a grid and take the measures one by one. The SQM-LU and SQM-LE devices should be linked to a computer by means of USB or ethernet connection respectively. These models are designed to be used at a monitoring station to take and register data continuously with the photometer fixed in one place.  All of them share the same characteristics. The Full Width at Half Maximum (FWHM) of the angular sensitivity is $\sim20^{\circ}$. The sensitivity to a point source $\sim19^{\circ}$ off-axis is a factor of 10 lower than on-axis. A point source $20^{\circ}$ and $40^{\circ}$ off-axis would register 3.0 and 5.0 astronomical magnitudes fainter, respectively (Unihedron SQM-L manual).

\subsection{Data acquisition procedure}
\subsubsection{Automatic acquisition}
The computer linked models of SQM photometers can also be used in the field with a laptop. Furthermore, when installed on top of a moving vehicle, they can provide readings of the sky brightness obtained during a trip. The SQM-L photometer needs a maximum integration time of about 20 seconds for the darkest skies in rural areas ($\sim21.9\;mag\;arcsec^{-2}$) to provide a reading. However the SQM-LU and SQM-LE models could be read every 5 seconds without lost of precision at these dark sites, since they provide the last measurement stored in its buffer. At this pace, one value each 100 m is obtained from a vehicle moving at a speed of 72 $km\;h^{-1}$. This procedure allows us to cover a wide area with enough spatial resolution.

To locate the position where each data point is taken we use a commodity GPS (Garmin eTrex Legend HCx) that registers the track at the same time. It is important to synchronize in time the computer and the GPS before each trip. At the end of a night of observations the data of the photometer and the GPS are joined to create a file with geographical position and sky brightness data for many points. To record the data the SQM-ReaderPro software was employed since it allows us to select the frequency of the acquisition. We have also used the RoadRunner software developed by Sociedad Malague\~na de Astronom\'ia that builds the file with photometric and positional data on the fly \citep{roadrunner}.

The procedure described above speeds up the data acquisition but it has some caveats. For a moving photometer the sky brightness obtained is a measure of the average values along the track during the integration time. However we do not expect an important change of sky brightness in such a small distance. Repeated tests of the same track at different speeds yield no significant differences in the values obtained. Given this invariance, we decide to match the speed to the requested spatial resolution of the final map. After several tests we settled for a frequency of 5 seconds interval.

\begin{figure}
  \includegraphics[width=90mm]{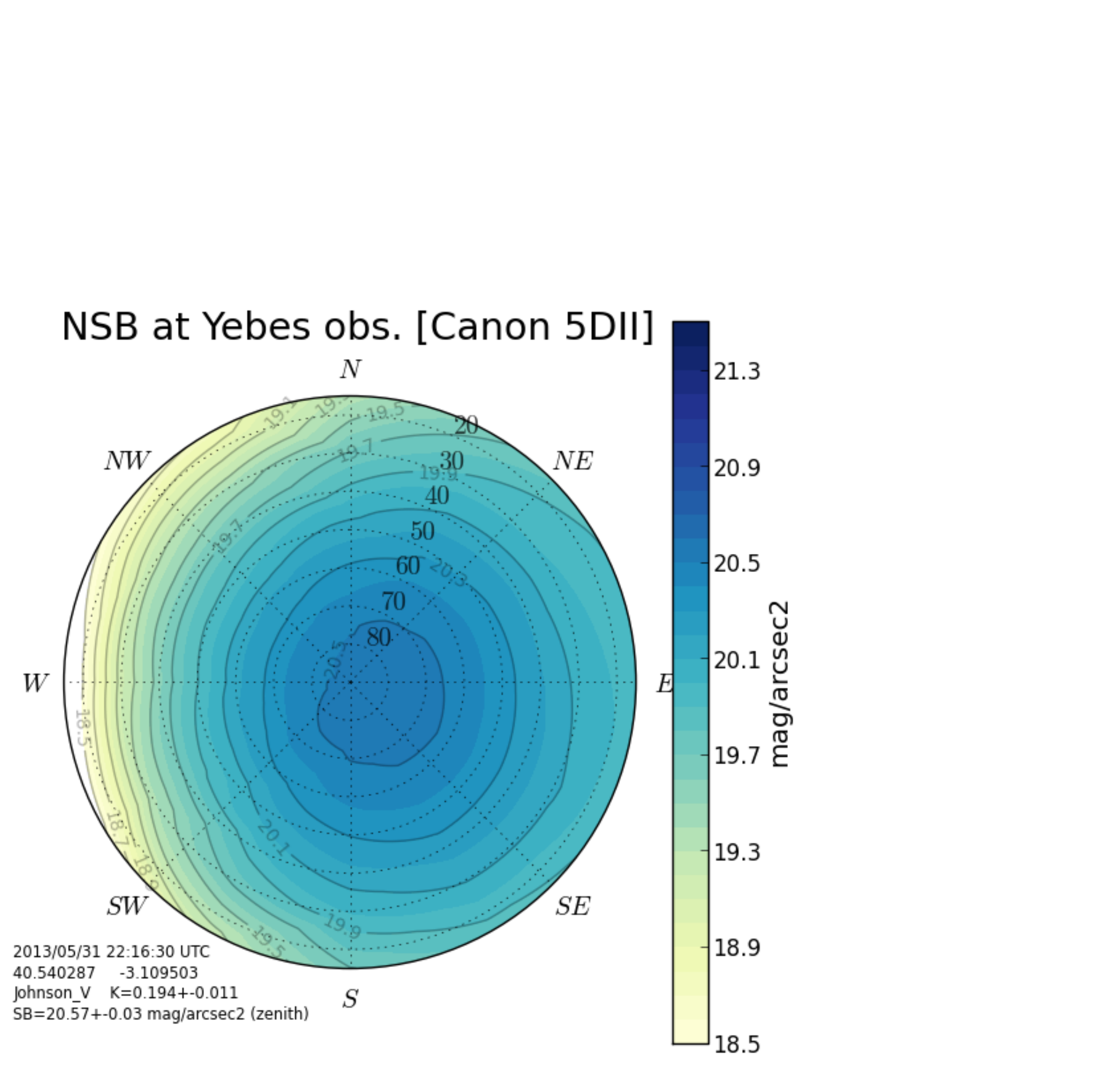}
  \caption{All-sky brightness map, in Johnson V band, obtained at
    Observatorio Astron\'omico de Yebes using a digital calibrated
    camera (DSLR). The asymmetry is due to the presence of Madrid to
    the West of the observatory. }
 \label{fig:Yebes_allsky}
\end{figure}

\begin{figure}
  \includegraphics[width=90mm]{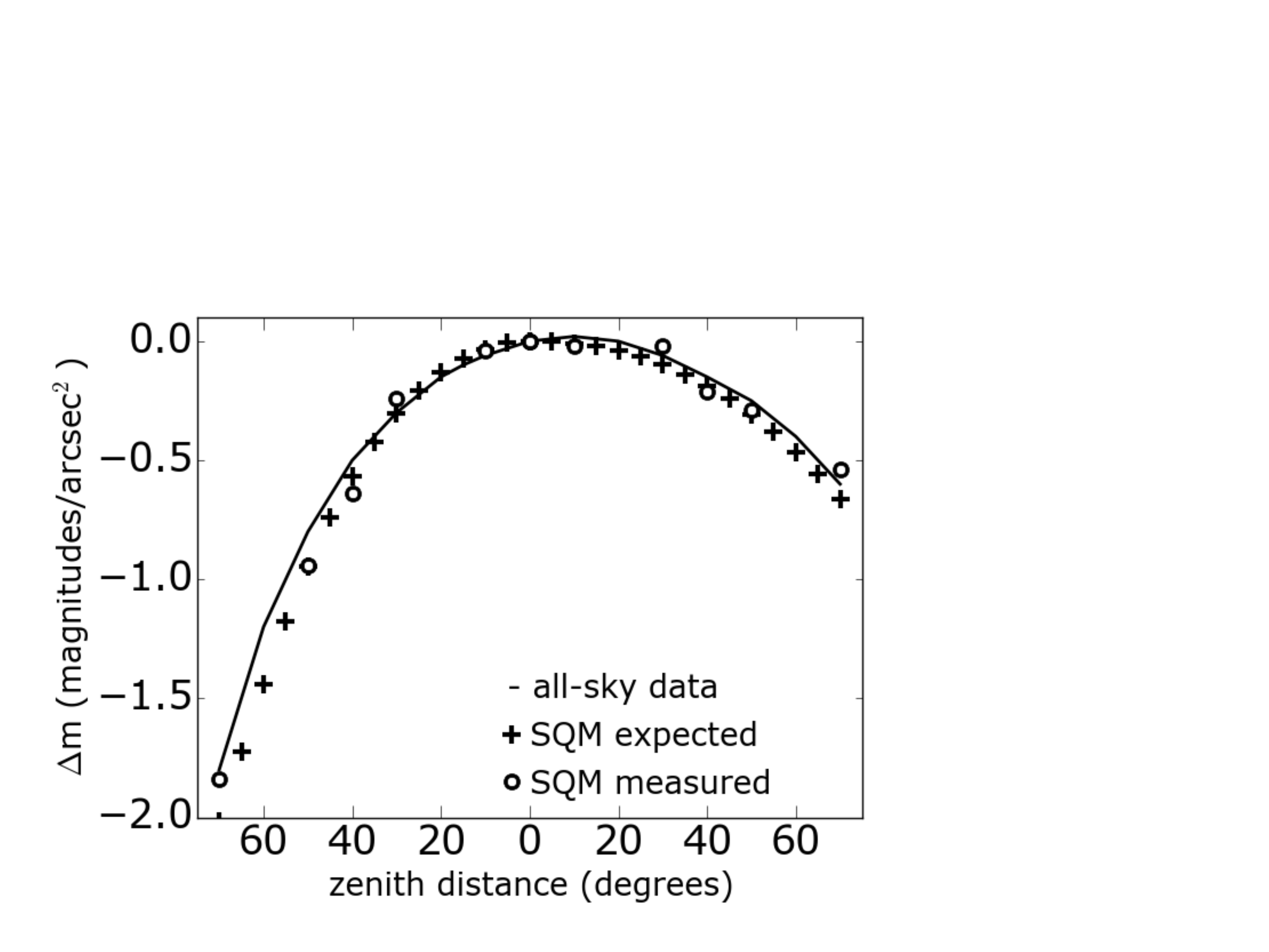}
  \caption{Differences in night sky brightness along NW-SE line derived from the all-sky
    map shown in fig \ref{fig:Yebes_allsky} (line). The plus sign and the open
    dots are the expected and measured NSB values using an SQM
    photometer pointed to the chosen directions, respectively.}
 \label{fig:Yebes_plot}
\end{figure}

\subsubsection[]{Night sky brightness at zenith}
The data presented in this work corresponds to sky brightness measured with the SQMs pointed at zenith. For this purpose the photometers were secured to the car top assuring that they were pointed perpendicular to the ground. The pointing precision is approximately 1 degree. However, when the vehicle is moving during a trip one can expect changes in the pointing due to the road inclination. A stabilizer suspension with a gimbal mount could be used to avoid completely this problem. Fortunately, roads are usually designed with low inclination. The highest slope in which we have obtained data has a  8.6\% inclination (4.9 degrees) in a small portion of the ascent to Puerto de Navacerrada (Sierra de Guadarrama mountain range, north of Madrid city). Furthermore, steep roads are usually located in dark places.

It should be borne in mind that the SQM-L photometer has an acceptance angle for the incoming light. Furthermore, there is a strong variation of the angular response of the photometer that drops with the angular distance to the optical axis. The SQM sky brightness measured when pointing the device towards the zenith is a convolution of the angular response with the radiance of the sky. The area of the sky that the photometer measures, increases with the angle from the optical axis.

It is possible to estimate the difference between the actual value of the night sky brightness at the precise point where the optical axis of the SQM-L points and the value measured with the photometer. The results of \cite{cinzano2005night} tests of the SQM photometer and a typical sky brightness dependency on the zenith distance \citep[][Table C.1]{patat2003ubvri}, who use the model of \cite{garstang1989night}) have been used. We found that the reading of the SQM-L should be slightly brighter than the actual value when the zenith angle increases (lower altitude) for a typical dark sky with an increase in brightness of 1 magnitude per square arcsec from zenith to horizon. We conclude that the SQM-L reading corresponding to an averaged sky region can be used as representative of the position where the photometer points. For typical polluted skies where the brightness near the horizon shows an important increase, these differences are relevant. An interesting result of this estimate is the small variation of night sky brightness near the zenith for natural dark areas. If the SQM-L is not correctly pointed to the zenith and it is aimed with an error of 10 degrees, the difference in measured value is less than 1\%.  The same is true for sites with moderate light pollution. 

To test this model calculation with real data obtained on the field, we obtained an all-sky image using a digital camera (DSLR) with a fisheye lens at Observatorio Astron\'omico de Yebes which is 70 km away from Madrid. The observations were calibrated using {\sc PyASB} open software \citep{nievas2013absolute,zamorano2015low}. The all-sky calibrated brightness map built with the processed picture shows an asymmetry, being the dark sky slightly off-zenith due to the light pollution coming from Madrid (located to the West) and the area around Alcal\'a de Henares (see Fig.~\ref{fig:Yebes_allsky}). Although the SQM band encompasses both Johnson {\it B} and {\it V} photometric bands, it is possible to use the all-sky data to predict SQM variation along one almucantar. We show in Fig.~\ref{fig:Yebes_plot} the variation of NSB with zenith angle (in relative units) measured along the NW-SE line for this map and the expected values after using the all-sky data and the angular response of the SQM. The expected and measured SQM values are brighter towards the horizon because the sky radiance increases when aiming the photometer from zenith to horizon.  Again the error due to inaccurate pointing around the zenith is of the same order and negligible.
\begin{figure}
  \includegraphics[width=90mm]{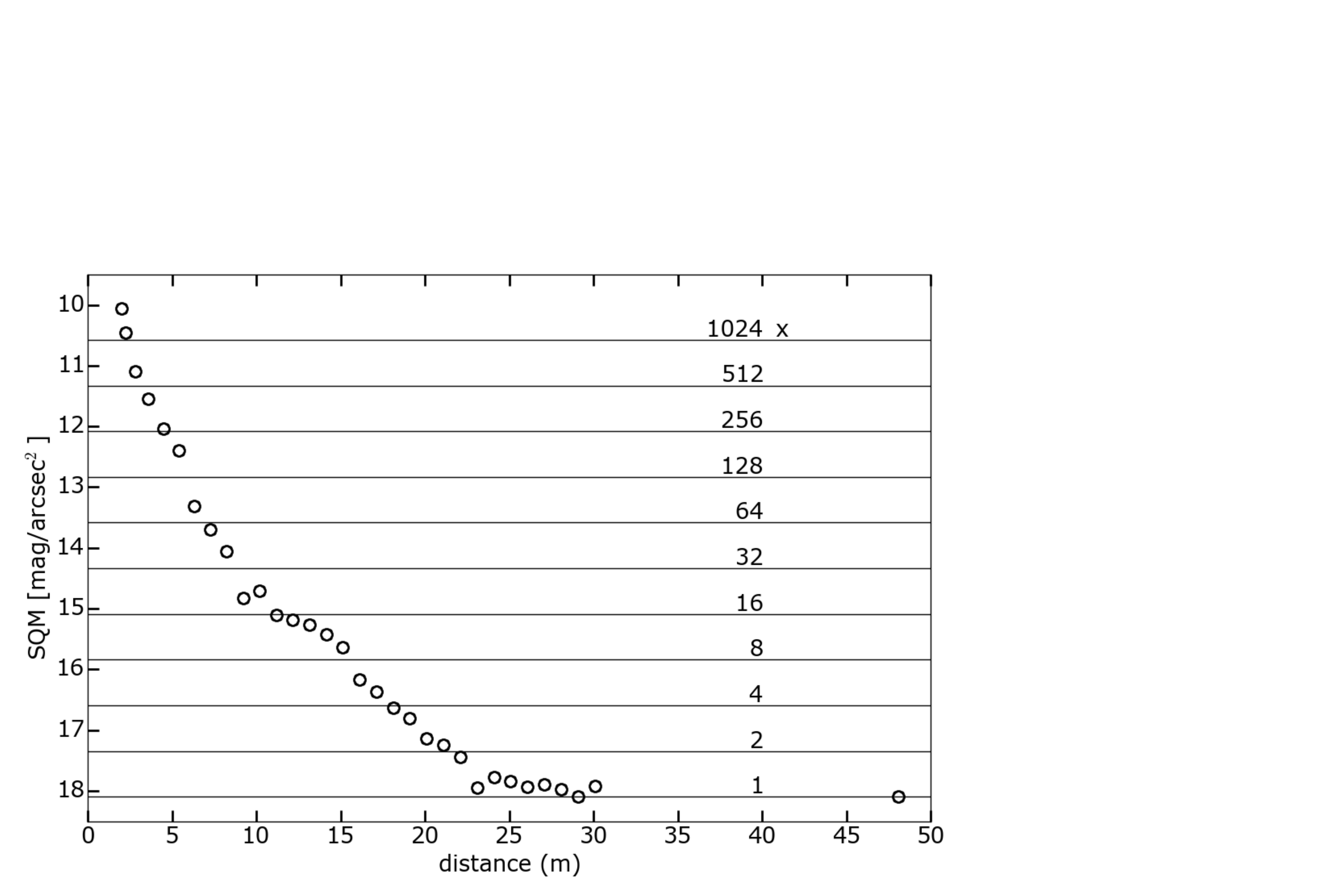}
  \caption{SQM values measured at different distances from an isolated lamppost pointing the photometer to the zenith (open circles). Measurements close than around 25 meters are brighter than the actual night sky brightness value by a factor indicated on the graph (labelled horizontal lines).}
 \label{fig:lamp_test}
\end{figure}

\subsubsection[]{Stray light}
The angular response of the SQM photometer has extended wings, with a drop of 100 times less sensitivity at 40 degrees from optical axis. However a bright artificial nearby source, such as a lamppost, could spoil the NSB measures. The cautious method should avoid illuminated roads and not to take measures inside populated areas. In this regard it is interesting to determine which is the safe distance where we could measure the sky brightness at zenith without contamination from a street lamp. Using a 9 m high lamppost with a sodium lamp (HPS), located at the end of a street in the boundary of a village, we obtained the actual value of the sky brightness ($18.2 \;mag\;arcsec^{-2}$) at $\sim 25\;m$ of distance (see Fig.~\ref{fig:lamp_test}). 

As expected, a series of peaks and valleys are recorded when driving along a main illuminated road.  Some experiments were performed to find whether it is possible to obtain actual sky brightness using the lower values measured. For instance the trips on 2010 May 16th and 17th included the northern part of the M-30 circle road around Madrid and the first 30 km of the E-5 and E-90 radial roads. We have found that the maximum speed that allows to have clean measures between lampposts separated 50m from each other is 60 $km\;h^{-1}$  when reading the photometer 5 times per second. This speed threshold is very low for a main road so we discarded this method. Nonetheless we used some data acquired in tracks of illuminated roads where the lamps were switched off due to maintenance or malfunction.

\begin{figure}
 \includegraphics[width=90mm]{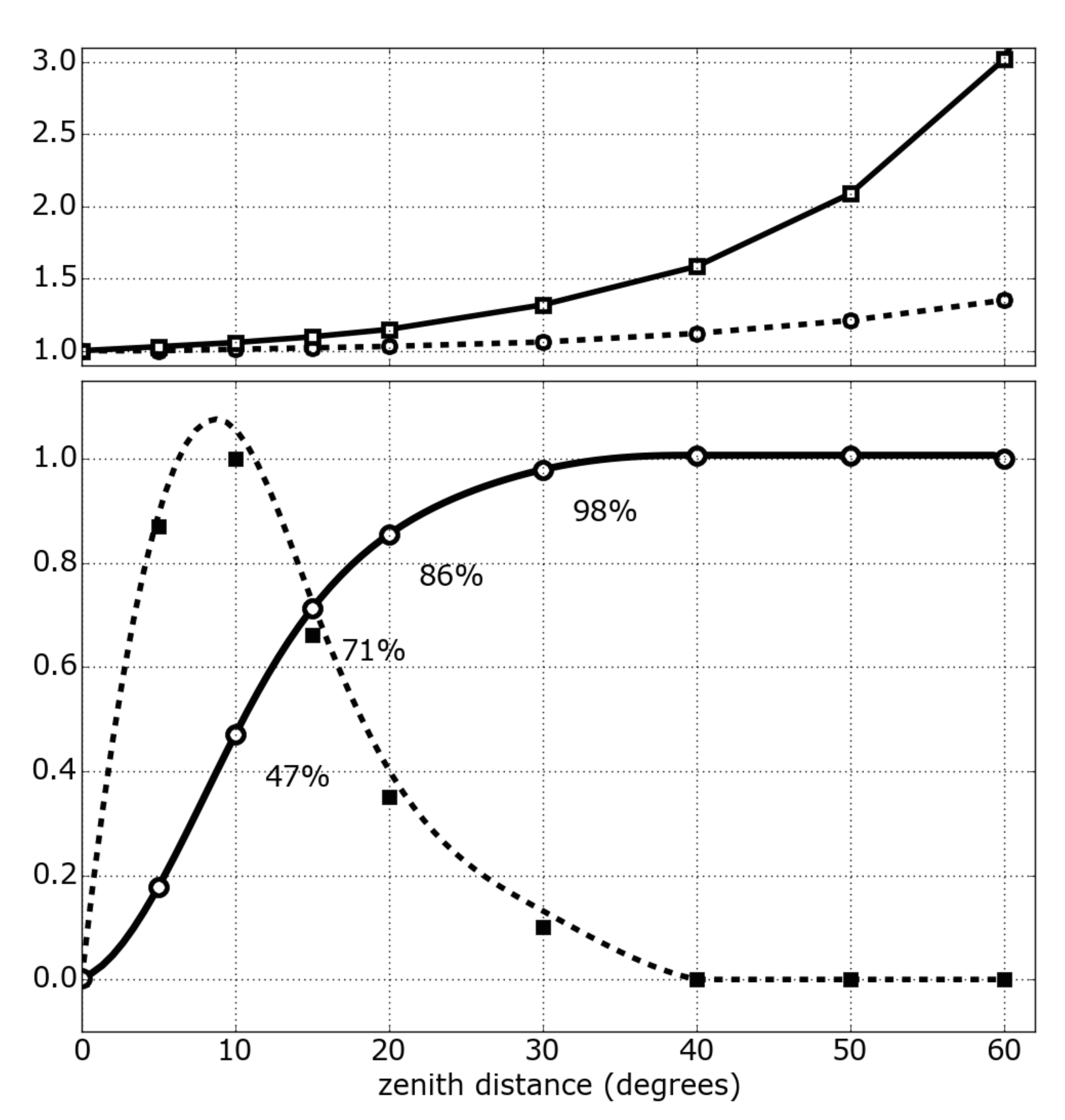}
 \caption{(Top) Nigh sky brightness variation from zenith to 30$^{\circ}$ height over the horizont for a very dark place in relative units (dashed line) and for Observatorio de Yebes (solid line). (Bottom) Cumulative radial response of a SQM-L photometer (solid line) along the zenith angle for a typical sky that brightens from zenith to horizon. The angular response of the photometer pointing to zenith has been weighted with the sky area that the photometer register for each zenith angle \citep[solid points, see][]{cinzano2005night}. The dashed line shows the signal registered on the photometer from different angles.} 
 \label{fig:SQMsint}
\end{figure}

In what follows we show that it is possible to use a simple diaphragm or baffle to mitigate the contamination from stray light. We have modeled the response of a SQM photometer, pointed to zenith, that is measuring the sky brightness at different scenarios. Fig.~\ref{fig:SQMsint} shows the night sky brightness variation (in relative linear units) with zenith distance for a dark place following \cite{patat2003ubvri} and the model of \citet{garstang1989night} and for the astronomical observatory of Yebes which is affected by the light pollution of Madrid in the W direction. The sky is around three times brighter at a zenith distance of 60$^{\circ}$ (30$^{\circ}$ above horizon) in the direction of the main pollution source.

The signal registered by the photometer for each direction can be obtained as $I(\theta,\phi)\;D(\theta,\phi)\;sin(\theta)\; d(\theta)\;d(\phi)$, being $D(\theta,\phi)$ the angular response. At this point it is important to note that the brightness measured by the SQM photometer will be the weighted average of the brightness at different angle distances from its optical axis, 

\begin{equation}
\bar{I} = \frac{\int_{0}^{2\pi} \int_{0}^{\pi/2} I(\theta,\phi)\;D(\theta,\phi)\;sin(\theta)\; d\theta\;d\phi}{\int_{0}^{2\pi} \int_{0}^{\pi/2} D(\theta,\phi)\;sin(\theta)\; d\theta\;d\phi}
\end{equation}

\noindent \citep[][eq. 1]{cinzano2005night}. From the cumulative response curve of the SQM we have found that the light measured with cylindrical baffles that block angles of 10, 20, 30 degrees from zenith are 47\%, 86\%, and 98\% respectively for a typical rural area.  We have not used any kind of baffle to prevent stray light from contaminating the measures. 

\subsubsection[]{Sources of wrong measures}
It is worth mentioning another sources of wrong or false data when using the automatic procedure. Some of them are evident, such as tunnels that are usually illuminated giving a false brighter measure. The same is true when the car passes under a bridge whose ceiling is illuminated by the car lights. Rural roads with trees at their sides are a problem when they are too narrow or there is a broad tree canopy covering the road. Brighter or dimmer data are obtained depending on the tree canopy being illuminated by the car lights or not in this kind of tracks. In the second case the branches of the trees can block the light from the sky. To obtain correct data during these tracks we decided to slow down or even to stop in places with open skies not obstructed by trees. 
In any case, a monotone variation of night sky brightness along the surveyed paths is expected and strong changes should be a sign of false data. Thus these unwanted data could be easily rejected after careful inspection or filtering of the data obtained in each series (see Section \ref{filtering}). It is therefore wise to keep a logbook and to take notes during the trips. For this reason, besides security matters, the minimum size of the team should be a driver and the person in charge of the data acquisition.

\subsubsection{Manual acquisition}

Measuring from a moving vehicle is a fast and efficient method to get NSB values but should be complemented with manual data acquisition near artificial lighting and inside the villages. One should find a place not contaminated with stray light to take careful uncontaminated readings with a manual SQM. Good locations are parks, rooftops, and even parking lots when their illumination is switched off after commercial hours. For main roads it is necessary to take detours to second or third order roads without lighting. The SQMs were coupled to a photographic tripod and leveled to the vertical (pointing to zenith) using a digital level. This allow us to achieve a pointing accuracy better than 1 degree, which is enough for our purposes.

\subsubsection{Campaigns}
The first data were acquired during two trainee projects performed by undergraduate students. Berenice Pila D\'iez obtained values at 3731 places during ten trips ranging from 55 to 242 km (April to May 2010) and with a total amount of 1146 km \citep{pila2010mapa}.  Alberto Fern\'andez made six trips (April to July 2011) with a total of 550 km. Some roads were done over to confirm the consistency of the measurements (see section \ref{repeatability}). Although the total area covered was enough to extract useful scientific conclusions we decided to perform some additional trips to less explored regions at the end of 2011. The analysis of the first campaigns provided many useful hints on the drawbacks of the method and on how to improve the procedure.

Manual acquisition was used to add points to those complicated areas were the automatic method could lead to false values. This is extremely important for the centre of Madrid community, where the big capital occupies an extended region with a radius of around 10 km.  During consecutive clear nights of new moon a total of 45 measures were taken in places without lights. Some readings were obtained at the roof of buildings. 

During the following years Carlos Tapia, Francisco Oca\~na, Jes\'us Gallego, Jaime Zamorano and Alejandro S\'anchez de Miguel performed additional campaigns that extended considerably the surveyed area. Our original aim was to measure in a dense mesh of locations around Madrid and to cover a wide range of sky brightness from the big city in the centre to the outer rural regions. The campaigns period extended until the time of writing this paper and continue. Considering only good nights when the data has not been rejected, the distance traveled while measuring is around 6753 km with over 50 thousand measures and 30,007 valid points. This data set covers an area of 5389 $km^2$ (assuming that the data points represent the night sky values of square cells of 2.2 $km^2$ side); most of them belong to {\em Comunidad de Madrid} region, with a total surface coverage of roughly 63\%. Some restricted access areas and mountain ranges were not surveyed. More details can be consulted at \cite{sanchezdemiguel2015variacion}.  

\subsection[]{Repeatability}\label{repeatability}
The data have been obtained during clear and moonless nights (first or last quarter moon and Moon below the horizon). A sky without clouds is mandatory because the clouds reflect light in polluted skies. We only used data of excellent quality nights. The campaigns were started on a priori good nights but the data of three of them were discarded at a later stage after checking the quality of the night (extinction and stability) using the NSB plots provided by fixed NSB monitor stations (see next section) and/or comparing with campaigns surveying the same areas. 

To control the quality of the night we used the data provided by the photometers of the astronomical observatory of Universidad Complutense de Madrid (Observatorio UCM, inside Madrid 40$^{\circ}$27${^\prime}$N, 03$^{\circ}$43.5${^\prime}$W,  IAU-MPC I86). The all sky camera (AstMon-UCM) monitors the astronomical quality of the sky at night including the sky brightness in the {\it B}, {\it V} and {\it R} Johnson photometric bands. See \citet{aceituno2011all}, for a description of AstMon and its capabilities. Thus we can determine the impact of clouds or aerosols on the sky brightness for this urban observatory located in the outskirts of Madrid city. The mean values of the sky brightness (since August 2010 to July 2011) are 19.20$\pm$0.12 and 17.90$\pm$0.09 for {\it B} and {\it V} respectively averaging only clear moonless nights. The increase in brightness for cloudy nights is $2.15\pm0.33 \;mag\;arcsec^{-2}$ and $3.69\pm0.68 \;mag\;arcsec^{-2}$, i.e. 7.2 and 30 times brighter respectively. Obviously measuring with cloudy skies (even partially covered) yields brighter and false values of the sky brightness. As expected, this effect is stronger in light polluted areas \citep{kyba2011cloud}.

\subsubsection[]{Variation of the NSB along the night}\label{night}
The night sky brightness on a location varies with the phase and altitude of the Moon, the season of the year, the hour of the night and the atmospheric conditions. An analysis of the sky brightness with cloudy skies performed with SQM photometers shows an increase of 0.3 magnitudes for 2 oktas cloudiness, i.e. 2/8 parts of the sky covered \citep{kyba2011cloud}.  It is therefore naive to think that a single measure is enough to get insight into the night sky brightness of a location. By selecting clear moonless nights we have discarded the effects of the moon and the reflecting clouds.

For fixed photometers used to monitor NSB, the temporal evolution along one night can be depicted with 'scotograms' \citep{puschnig2014night}. The comparison of these graphs with the typical plots for a good quality night provides the information to reject the data taken during some nights.  Besides,  there is a variation along the night in the sense that the big cities have brighter skies at the beginning of the night while the second part of the night is darker \citep[see for instance][]{bara2015report}. This is the result of a lesser activity of the city and the switch of ornamental lights. For Madrid city (Observatorio UCM station) the typical evolution for a perfectly clear and moonless night shows a difference of $0.2\;mag\;arcsec^{-2}$ \citep{sanchezdemiguel2015variacion}.  

We expect to find darker skies at the end of typical nights for all the surveyed areas although the effect is diluted as one goes farther away from Madrid city centre. Madrid is a big urban area and the effect of its light pollution can be detected at large distances. In fact the all-sky images obtained at 130 km shows its light glow. At this location (Villaverde del Ducado monitor station) there is no difference between NSB at zenith between the beginning and end of the night except for the effect of the Milky Way. 

This paper is devoted to the spatial variation of the night sky brightness in a wide area. Our method does not allow to obtain at the same time the temporal evolution. The study of both spatial and temporal evolution requires a network of fixed monitor stations, such as the one established in Hong Kong area \citep{pun2014contributions}, which is now under development around and inside Madrid. Besides Observatorio UCM, during most of the campaigns the monitor stations near Alcal\'a de Henares (40$^{\circ}$26${^\prime}$N 03$^{\circ}$18${^\prime}$W), at Observatorio de Yebes (40$^{\circ}$31.5$^{\prime}$N 3$^{\circ}$05.5${^\prime}$W, IAU MPC 491) and Villaverde del Ducado (41$^{\circ}$00${^\prime}$N 2$^{\circ}$29.5${^\prime}$W), at 40, 70 and 130 km from Madrid respectively, were operative. For our purposes it is now enough to remember  that whithin Madrid the maximum difference in the NSB is $0.2\;mag\;arcsec^{-2}$ along the night for a clear and moonless night. No attempt has been made to correct for time of the night.

\subsubsection[]{Intercomparison field test}\label{field}
\begin{figure}
\includegraphics[width=120mm]{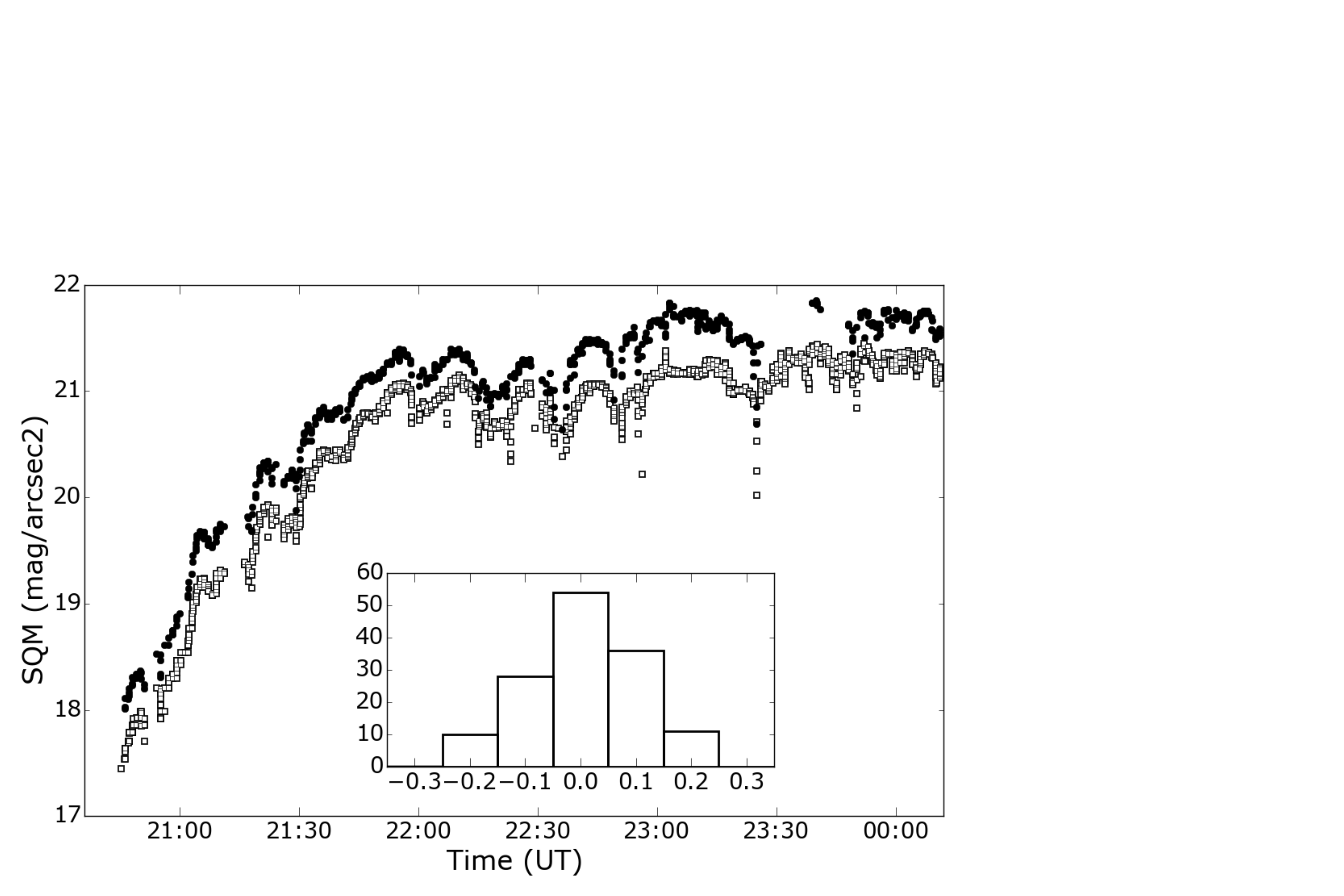}
\caption{NSB at zenith during a test campaign comparing two photometers. Open and filled points corresponds to a simultaneous measure during the night of 2014-01-25 with two SQM photometers (\#1716 \& \#1738). The measures from \#1738 have been shifted upward by $+0.4\; mag\;arcsec^{-2}$ for better visualization. The differences are plotted in the histogram.}
\label{fig:Tarancon}
\end{figure}

Areas near the big cities with a high degree of light pollution are
difficult places to obtain actual values of the night sky brightness  ($\sim18\;mag\;arcsec^{-2}$). A misalignment of the photometer with the zenith could yield incorrect results. To perform a repeatability and inter-comparison test, the two photometers most used for this survey (serial no. \#1716 \& \#1738) were independently secured on top of a car traveling from and to Madrid. The test was performed on the night 2014 January 25-26 during a trip from Madrid to Oca\~na, Mota del Cuervo, La Almarcha (160 km from Madrid city centre), and Taranc\'on. The complete track (275 km) was surveyed in 3h15m. The simultaneous data are shown in Fig.~\ref{fig:Tarancon}. We found an offset up to 0.2 $mag\;arcsec^{-2}$ between the values of zenith night sky brightness measured by the two photometers. It is interesting to note that this difference varies along the track. Our conclusion is that there was a difference in pointing between the photometers in the sense that one of them pointed slightly forward while the other one pointed slightly backwards. This is why the first one measures brighter values when traveling towards a village and dimmer values when leaving with respect to the other one. Another less likely explanation for this difference could be the asymmetrical azimuth response of the SQM since the photometers were placed one perpendicular to the other. 

The differences obtained should be considered as upper limits since the night of test was not completely clear and some dim clouds or aerosols were present. The sky brightness measured on a different trip on the night of 2013 December 29, with 40 km of coincidence, is $\sim 0.2\;mag\;arcsec^{-2}$ brighter. The aerosol content (measured during the day, \cite{holben1998aeronet}) was low for the two nights, 0.051 and 0.020 aerosol optical depth respectively. We conclude than the differences caused by misalignment or orientation of the photometers are lower that those due to differences in the quality of the night. 

\subsubsection[]{Photometer cross-calibration}
\begin{figure}
\includegraphics[width=120mm]{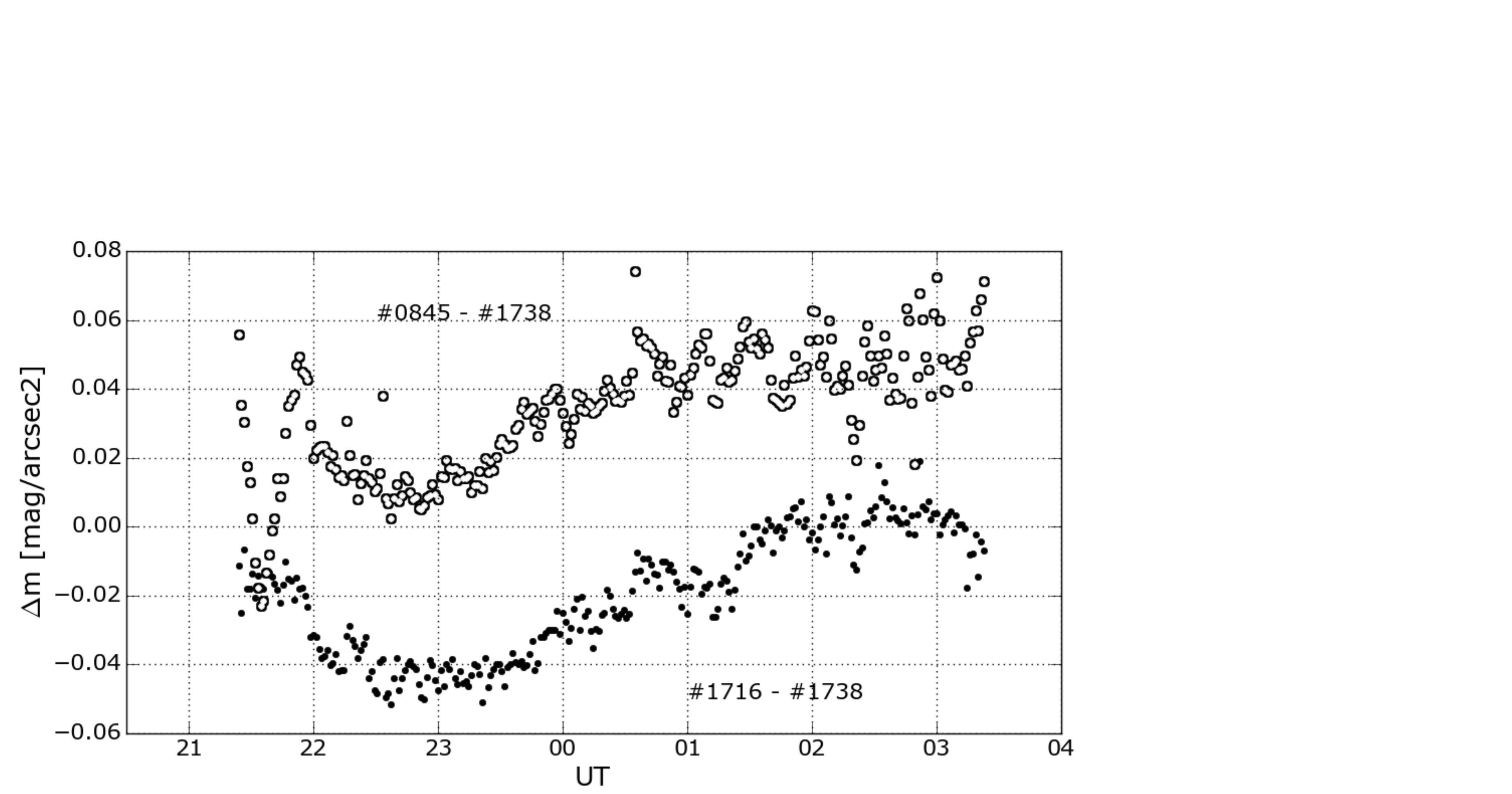}
\caption{Pairwise differences along the night of 2015-07-18 for the three SQM devices used in this work. The photometers were set fixed in a rural area pointing to the zenith.}\label{fig:offsets}
\end{figure}

SQM photometers have been calibrated at the factory and the manufacturer claims a precision and differences in zero point between devices of 0.1 magnitudes as long as some cautions are taken, as stated on the manual of the photometers. This is enough for the purposes of this research bearing in mind the other sources of error. When using different photometers it is advisable to check the zero point differences using a simple setup as that built to determine differences for the twelve photometers used in the NixNox Project \citep{zamorano2011proyecto,zamorano2015low}. Most of the measures (99.1\%) were made with the SQM devices with serial numbers \#845, \#1716 and \#1738 (1847, 18857 \& 9026 valid points respectively). Some inter-comparison tests were performed between these devices by means of the above mentioned test measuring from a moving vehicle or those performed during the 2014 LoNNe Inter-comparison Campaign (\cite{bara2015report}). We show in Fig.~\ref{fig:offsets} the resulting offsets corrections for these three SQM devices after a test with simultaneous measures during a night with NSB zenith values in a wide range from twilight ($17\;mag\;arcsec^{-2}$) to dark night ($21.2\;mag\;arcsec^{-2}$) in a rural area. The offset corrections found are listed in Table~\ref{tab:log_book}. These corrections were not applied to each photometer data to put all the values in the same reference frame. Given that these zero offsets are much smaller than the other sources of error (see sections \ref{night} and \ref{field}), no attempt has been made to perform an absolute calibration.

\begin{table}
      \centering
      \caption[]{Offset for most used SQM photometers}
         \label{tab:log_book}
\begin{tabular}{l  l  l  c}     % 5 columns 
\hline \hline
\# & Serial & Type & offset \\
\hline       
1 & 0845 & SQM-LE & $+0.03\pm$0.02\\
2 & 1716 & SQM-LU & $-0.02\pm$0.02\\
3 & 1738 & SQM-LU & reference \\
\hline
\end{tabular}
\end{table}

\begin{table}
      \centering
      \caption[]{Distribution of NSB dispersion in cells }
         \label{tab:dispersion}
\begin{tabular}{l  r  r }     % 3 columns 
\hline \hline
 & \multicolumn{2}{l}{Cells width (arcsec)}\\
dispersion  & 30 & 60  \\
\hline 
$\sigma < $ 0.1       & 80\% & 75\% \\
0.1 $> \sigma <$ 0.2   & 17\% & 22\% \\
$\sigma >$ 0.2         & 3\% & 3.7\% \\
\hline
\end{tabular}
\end{table}

Nonetheless we have summed up all the effects that we can not control and we assign an accuracy of $\sigma=0.1 \;mag\;arcsec^{-2}$ for the resulting map. We will come back to this calculation in the next section.

\section[]{Results}
\subsection[]{Map of the night sky brightness}
\subsubsection[]{Filtering the NSB values}\label{filtering}

The bulk of the data contains measures that should be rejected since they have been contaminated with stray light from luminaries, lights from gas stations, car lights and even light from our own vehicle and reflected on trees, to name but a few. Given that a monotone variation of the NSB at zenith is expected it is in theory possible to filter out outliers using an automatic procedure. However, after some unsuccessful tries, we preferred to use a manual method in which we made a careful supervised inspection of every data point.

The procedure is as follows. First of all we create a Keyhole Markup Language (KML) file to visualize the data with Google Earth. Our files contain the name of the observer, the SQM serial number, the time, the coordinates and the NSB value. Browsing with Google Earth along the tracks we detect wrong data (generally too bright) that we flag to be rejected. Using the geographical information it is easy to find an explanation for bright values. After that we build another KML file that contains the result of averaging the values in cells 15 or 30 arcsec wide. The cells are color coded using the NPS scale \citep{duriscoe2014relation}. A second pass through the data with the help of the cells help us determine more subtle wrong data that were not filtered out first  because we do not expect abrupt changes between consecutive cells. Instead of averaging NSB magnitudes (logaritmic scale) we prefer to determine the mean value inside a cell using a linear scale. The outlier measurements displace the mean to values too bright when comparing to adjacent cells.  This method is very useful to detect wrong data points. The process is repeated until convergence, usually in only two steps.

\begin{figure}
 \includegraphics[width=120mm]{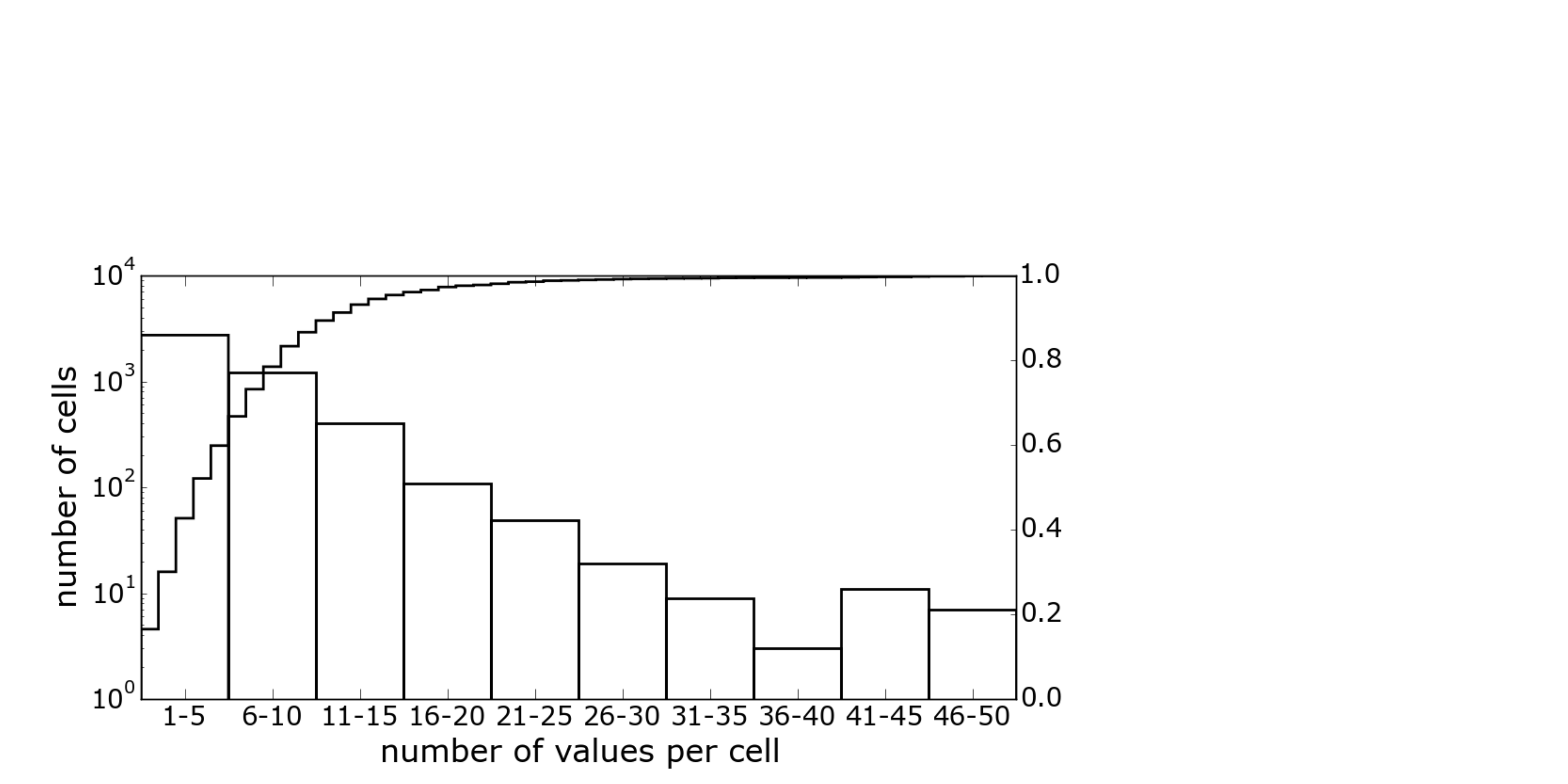}
 \caption{Distribution of cells with different number of valid NSB data points inside. The bin size has been set to 5 and the scale is logarithmic. The line represents the cumulative histogram (right Y axis).}
 \label{fig:Points_Hist}
\end{figure}

The method would not work properly when the density of values is low. This is why one of us is driving slowly while another person supervises the results during the data acquisition. The statistics of valid data points per cell is plotted in Fig.~\ref{fig:Points_Hist}. Most of the 4587 cells have a small number of NSB data with 1373 cells with  30\%  of the cells containing only 1 or 2 points inside, and cells with more than 10 points summing only for 13\%.   

\begin{figure}
 \includegraphics[width=120mm]{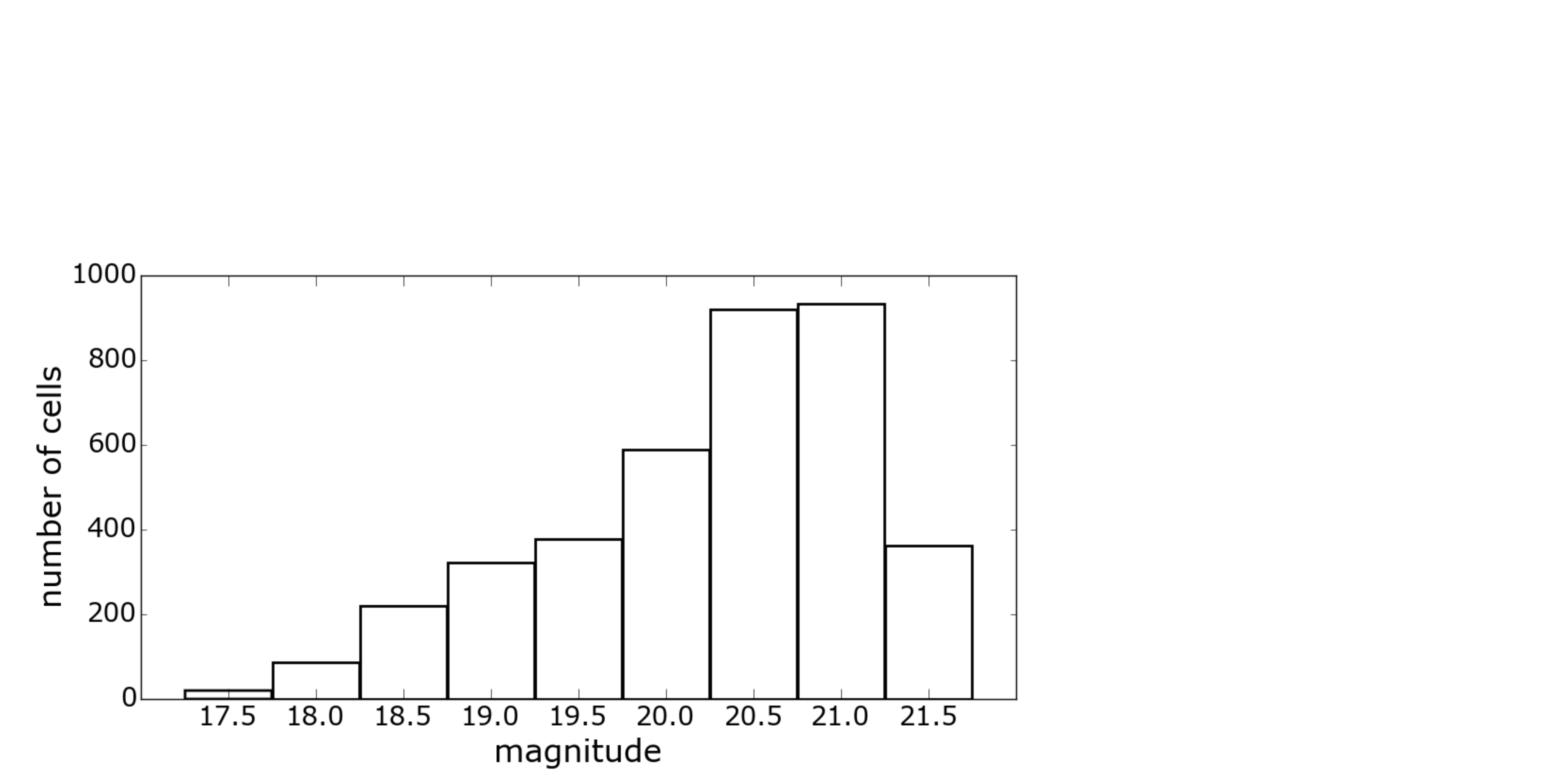}
 \caption{Histogram that represents the distribution of cells according to the NSB. Most of the cells with values around $17.5\; mag\;arcsec^{-2}$ are close to Madrid.}
 \label{fig:Cells_Hist}
\end{figure}

The final map is made with the above mentioned cells, i.e. the spatial resolution is that of the selected cell size. As we wish to compare with satellite data the position of the cells has been matched to those of the data products provided by NOAO (see section \ref{satellite}). Cells of 30 arcsec side correspond to rectangular areas of 0.71 km $\times$ 0.92 km at the latitude of Madrid.
 
To analyze the resulting map we have derived the dispersion of the measures inside each cell. We found that 80\% of the cells have $\sigma< 0.1\;mag\;arcsec^{-2}$, while only 3\% have values higher than $\sigma>0.2\;mag\;arcsec^{-2}$ (see Table~\ref{tab:dispersion}). This result supports our claim to adopt $0.1\;mag\;arcsec^{-2}$ as the internal accuracy rms of the map.

\begin{figure}
 \includegraphics[width=\textwidth]{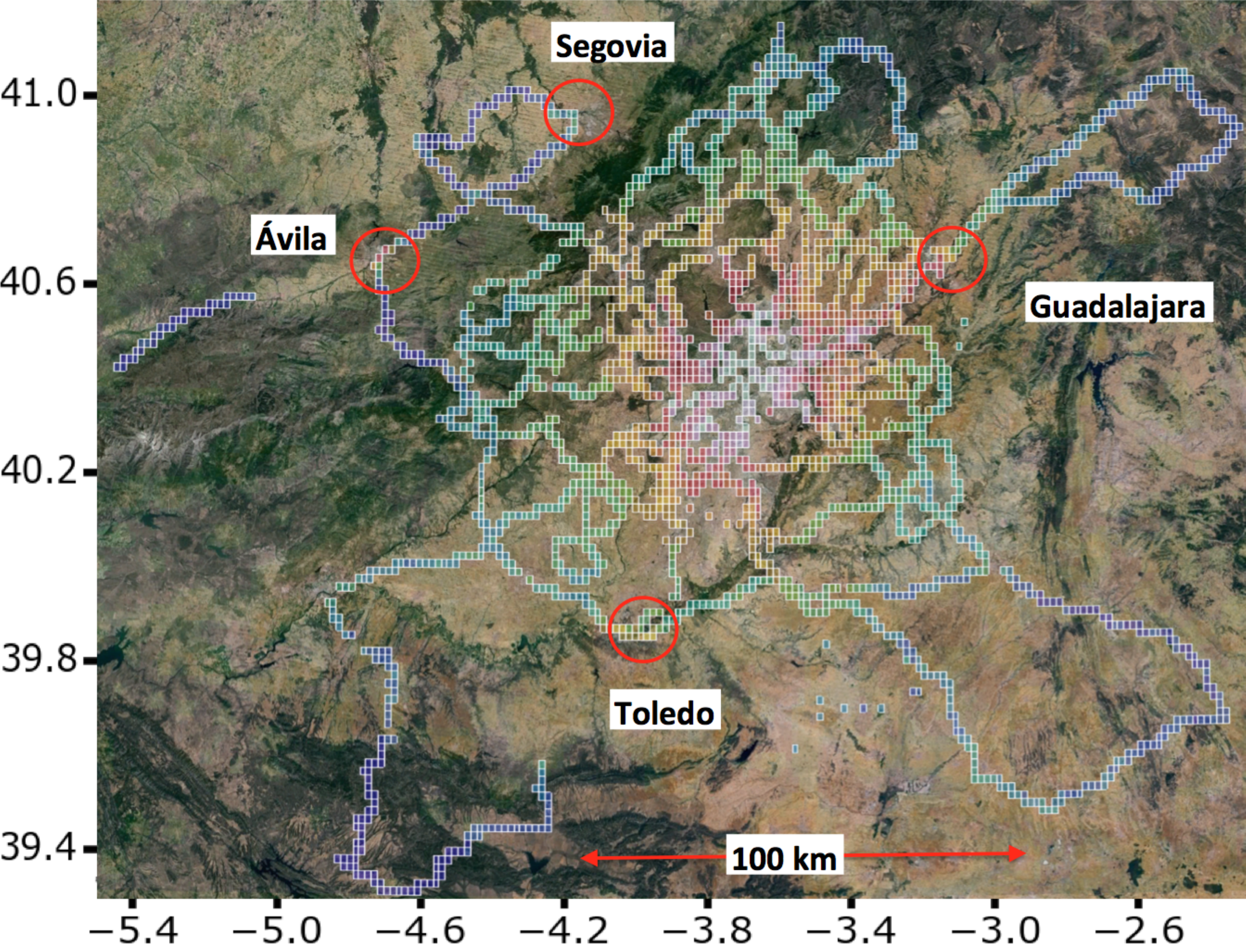}
 \caption{Night Sky Brightness map of the centre of the Iberian peninsula using cells 60 arcsec wide. The map has been color coded with the US NPS color table and plotted using Google Earth. The circles mark some cities around Madrid. }
 \label{fig:SQM_CAM}
\end{figure}

Fig.~\ref{fig:Cells_Hist} shows the distribution of cells according to the magnitude of the NSB. There are surveyed cells in all the range of sky brightness from the very bright inside Madrid ($17.0\;mag\;arcsec^{-2}$) to the rural areas with almost unpolluted skies and near the natural sky brightness ($\sim21.5\;mag\;arcsec^{-2}$).

\subsubsection[]{Radial variation of NSB}\label{Radial}

\begin{figure}
 \includegraphics[width=90mm]{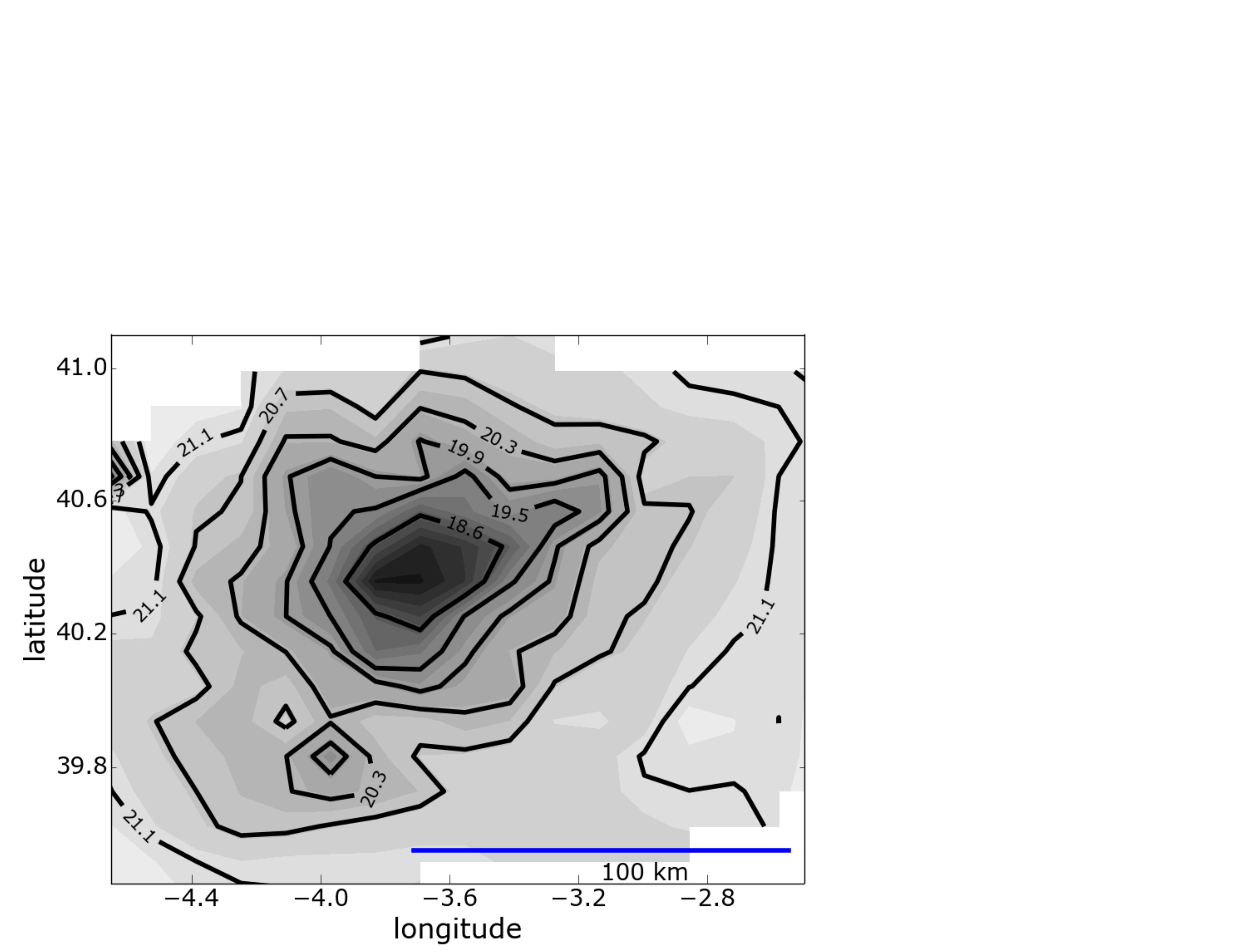}
 \caption{Contour plot of the NSB around Madrid built by linear interpolation of the cells values. The map covers the central part of Figure \ref{fig:SQM_CAM}. See text for details.}
 \label{fig:Contour}
\end{figure}

Madrid city (at 650 m altitude) and most of the region ({\em Comunidad de Madrid}) lies on the central highland plateau of the Iberian Peninsula. The western limit is defined by the Guadarrama mountain range. The population of {\em Comunidad de Madrid} is around 6.5 millions of inhabitants, most of them living in Madrid city ($\sim3.2$ millions). The metropolitan area of Madrid (population of 5.4 millions of inhabitans inside the circle of 27 km centered in Madrid) is very extended and asymmetric. In particular there are extensions along the main roads that are easily detected on the nocturnal satellite images (see Figure~\ref{fig:DMSP_CAM}). The most prominent towards the NE is the link between Madrid and Guadalajara along the Henares corridor. 

Fig.~\ref{fig:Contour} shows a contour plot of the NSB that has been built using the values of the 60 arcsec wide cells. A linear interpolation has been used to fill the cells without data. Since there are many gaps this model map should be used only for qualitative purposes. The isophotes are far from being round circles around Madrid city centre and they are found to be in good agreement with satellite images in places near Madrid. At longer distances the effects of Toledo (39.8N,4.0E) and \'Avila (40.7N,4.7E) are evident.

\begin{figure}
 \includegraphics[width=120mm]{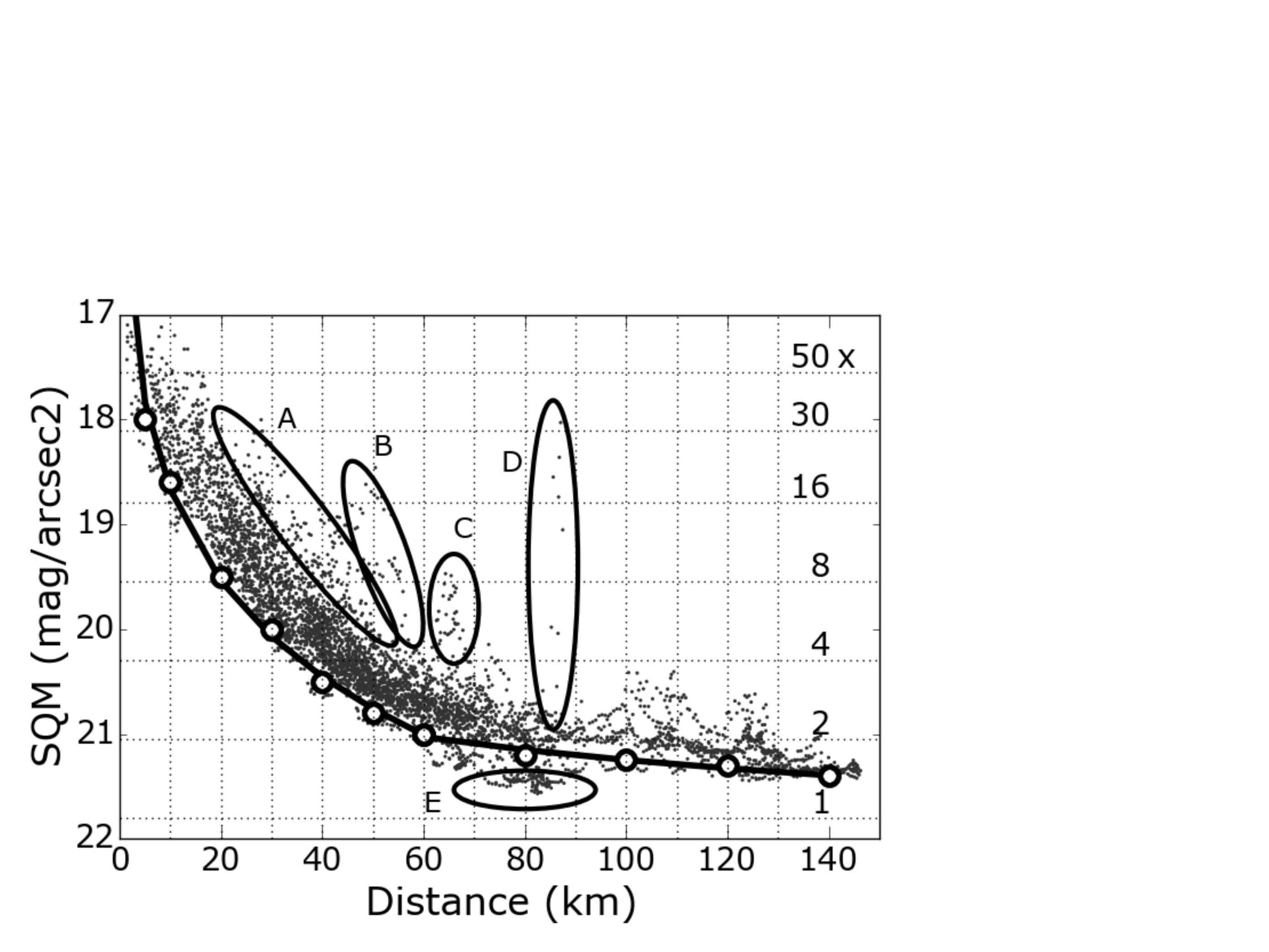}
 \caption{Radial profile of NSB around Madrid using the data averaged in cell 30 arcsec wide. Some data points have been marked: (A) Henares corridor from Madrid to Guadalajara (B); (C) belong to cells around Toledo and (D) in the neighborhood of \'Avila; the points marked as (E) belong to the NW of Madrid at the other side of the Guadarrama mountain range. The open points follows the lower bound of NSB distribution along the distance and the lines are exponential fits to these eye estimates. The numbers at right indicates the relative brightness (linear scale) with respect to a natural sky with SQM = 21.8 $mag\;arcsec^{-2}$. See text for explanation.}
 \label{fig:Radial_2}
\end{figure}

Some tracks were designed to study the radial dependence of the NSB with distance. They maximum distances reached are 145 km (towards W and SW) and 130 and 140 km for NE and SE directions (see Fig.~\ref{fig:SQM_CAM}). The distance from Madrid city centre (40.4N, 3.7E) to the locations where specific values of the NSB are reached (defining an isophote) depends on the direction. Therefore it is not easy to find a radial dependence from Madrid city (the primary light pollution source) using an azimuth average profile. This is illustrated in Fig.~\ref{fig:Radial_2}, a radial plot that has been built with the data of the 30 arcsec cells and using the distance of the centre of each cell to Madrid city centre. The expected trend for darker skies at longer distances from Madrid is found with a dispersion due to the spotted distribution of secondary light pollution sources all over the surveyed area and other effects previously mentioned. However some data points clearly deviate from the main trend. These data points belong to the corridor that links Madrid and Guadalajara (80,000 inhabitants) and to the areas around Toledo (80,000) and \'Avila (60,000). It is interesting to note some points, darker than expected, that correspond to an area in NW past the mountain range. The open points marks the bound defined by the darkest values of NSB at selected distances which have been fitted with exponential functions.

Due to the intrinsic and wide dispersion it makes no sense to fit a function to the median values of the distribution. We have estimated the upper bound of the NSB values for several distances from Madrid city centre. To highlight these values, two exponential functions have been fitted with exponents -1.0 for distances up to 60 km and another one with a lower slope of -0.4 for greater distances. The exponential functions have been fitted to NSB in linear scale trying to recover the -2.5 exponent of the Walker Law  $I = a\; P\; d^{-2.5}$ where $P$ is population and $d$ distance \citep{walker1977effects}.

\subsection[]{Comparison with satellite data}\label{satellite}

The satellites that take images of the Earth at night from space provide radiance data of the light emitted upwards, which is the main source of light pollution. The models for the dispersion of the light by the atmosphere should link this data with the brightness of the sky observed from ground.  \cite{cinzano2000artificial} and \cite{cinzano2004night} used the radiance data provided by the Defense Meteorological Satellite Program ({\it DMSP}) Operational Linescan System ({\it OLS}) to model the night sky brightness caused by artificial lights around the world. 

\begin{figure}
 \includegraphics[width=\textwidth]{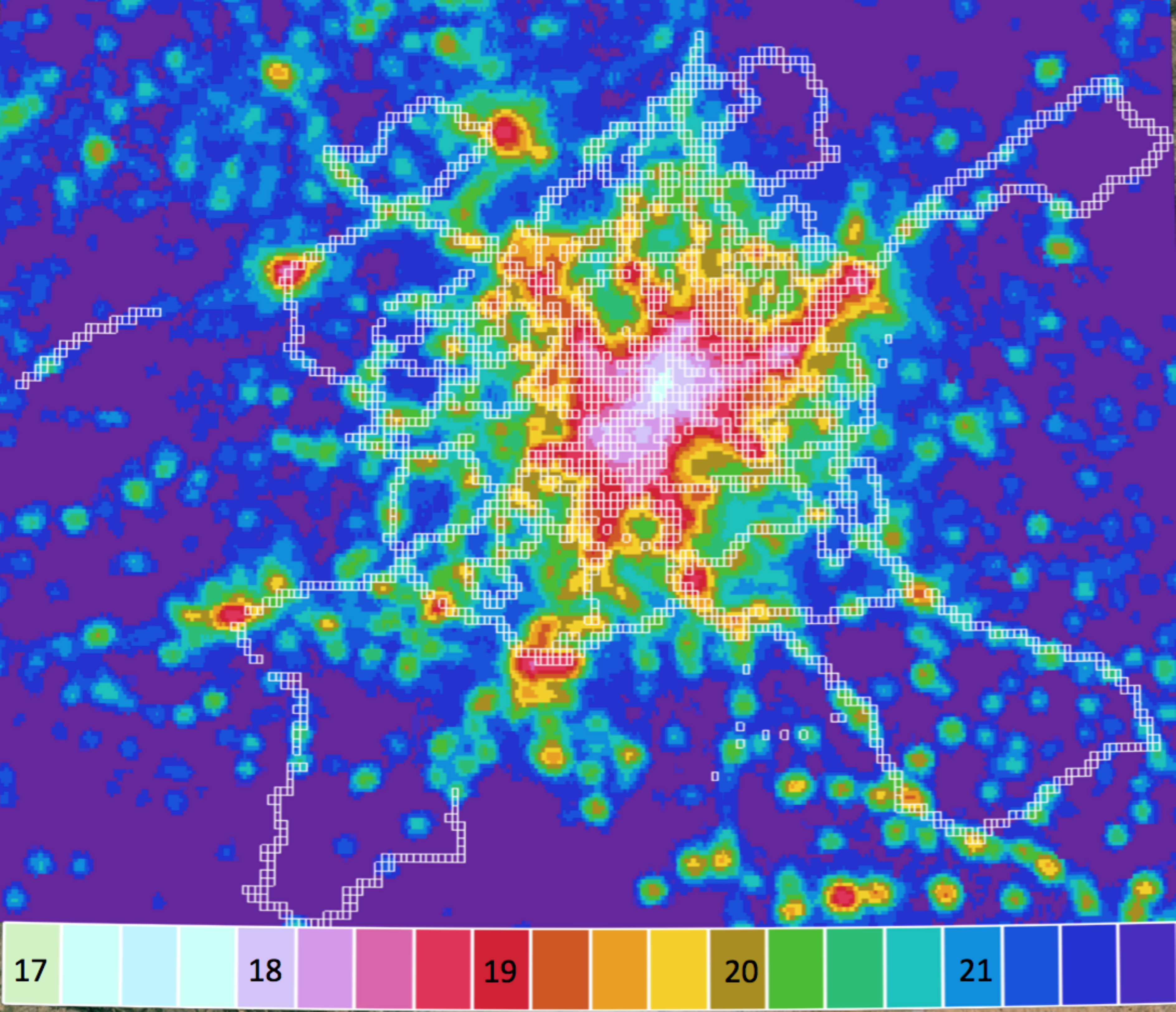}
 \caption{{\it DMSP/OLS} scaled radiance map of the centre of the Iberian peninsula using cells 30 arcsec wide. The white rectangles show the position where there are measured values of the NSB. The area displayed is 268 $\times$ 210 km$^2$. The radiance values have been transformed to expected NSB brightness using the fit found in Section \ref{satellite} (see Figure \ref{fig:SQM_DMSP}).}
 \label{fig:DMSP_CAM}
\end{figure}

The {\it DMSP-OLS} has two bands visible (VIS) and thermal-infrarred (TIR) designed to map clouds both at day and at night. We are interested in the visible band which detects photons in the 580 to 910 nm band even during the night. The detector is a photomultiplier tube (PMT) whose gain could be adapted according to the radiance. Although {\it DMSP-OLS} provides daily global coverage, the National Geophysical Data Center (NGDC) produces annual global cloud-free nighttime lights data sets, named {\it OLS Stable Lights products}, which are representative of clear and moonless nights \citep{elvidge1997mapping}. Details on the calibration can be read in \cite{hsu2015dmsp}. To summarize, data obtained at different gains are combined and related to radiances based on the pre-flights sensor calibration to provide global radiance calibrated images.

The spatial resolution of the global map is defined by the telescope pixel: 0.55 km at high resolution (fine mode). The final products have the information in cells of $30 \times 30 \;arcsec^2$. Since we made the map of the NSB using the same mesh of cells, i.e. same size and centre of cells, the comparison is performed with data belonging to the same area and position both from ground and space measures. The radiance map presented in Fig.~\ref{fig:DMSP_CAM} is composed of 379$\times$266 cells. The nocturnal radiances in it correspond to the map generated by \cite{ hsu2015dmsp} based on the data of the {\it DMSP/OLS} imagery of 2011. The corresponding cells with NSB data have been highlighted for comparison.

\begin{figure}
 \includegraphics[width=120mm]{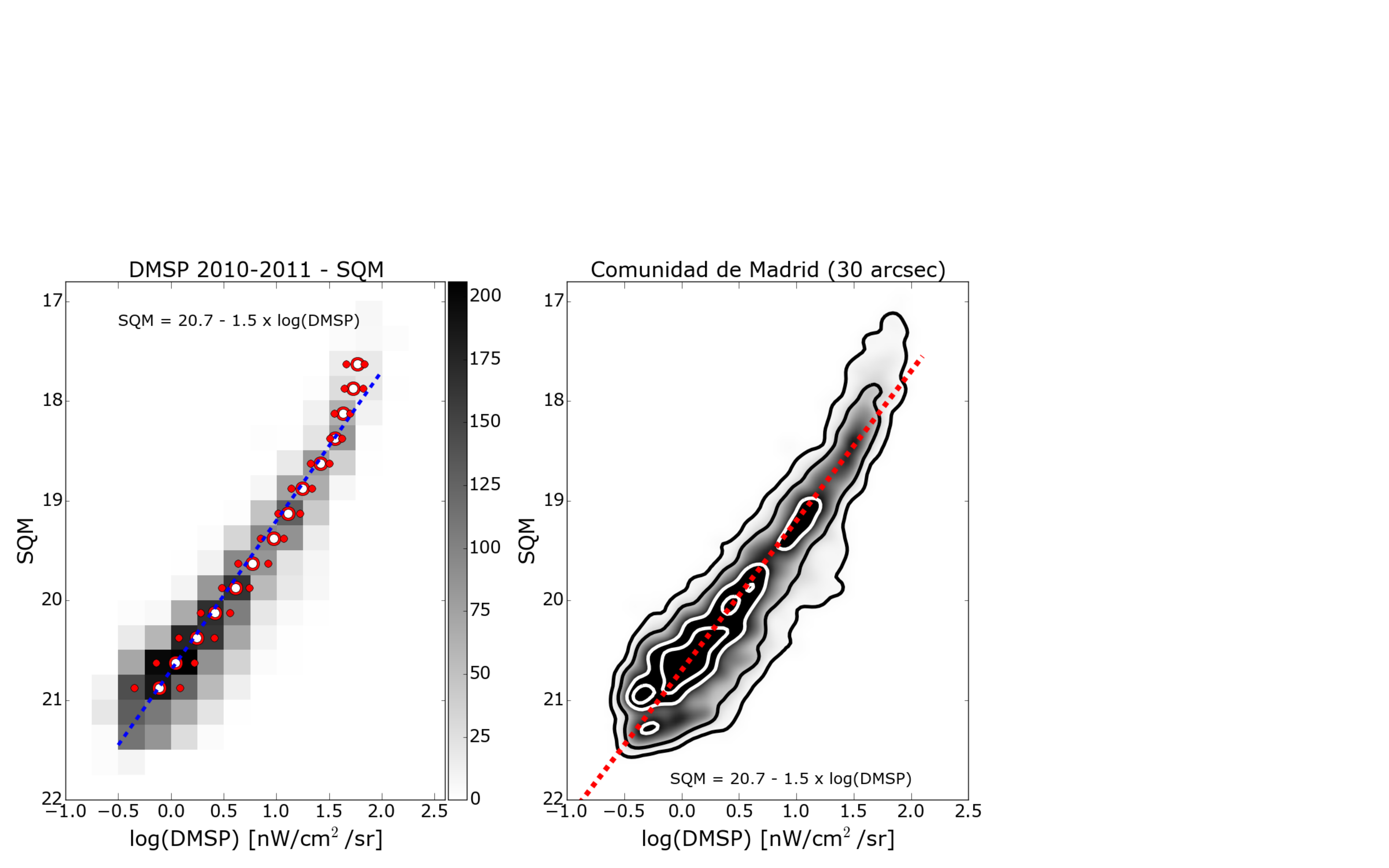}
 \caption{Relationship between the NSB values and the radiances measured by {\it DMSP/OLS} using 30 arcsec wide cells. The left panel shows a two dimensional histogram of the data binned in steps of 0.25 (in log of radiance [$nW\;cm^{-2}\;sr^{-1}$]) by 0.2 in $mag\;arcsec^{-2}$. The red point mark the median and the 25\% and 75\% percentiles. In addition, the linear fit has been overplotted in blue. The right panel shows the same data in a density plot with the same linear trend plotted in red. }
 \label{fig:SQM_DMSP}
\end{figure}

The results of the comparison between nocturnal satellite radiance data and the night sky brightness measured from the ground is displayed in Fig.~\ref{fig:SQM_DMSP}. In order to highlight the close relationship, a two dimensional histogram has been built (left panel) with the data binned in steps of 0.25 in log of radiance [$nW\;cm^{-2}\;sr^{-1}$] by 0.2 in NSB [$mag\;arcsec^{-2}$]. The number of 30 arcsec cells wide in each bin has been colour-coded in a gray scale. Additional points mark the median values of the NSB in each bin. The linear fit found is NSB([SQM] =  20.7 - 1.5$\;$log$_{10}$({\it DMSP}[$nW\;cm^{-2}\;sr^{-1}$]). We have removed from the fit a small number a small number of bright cells that do not follow the relationship. The same result is presented on the right panel as a density plot.

\subsection[]{Comparison with similar studies}\label{comparison}
This is the first published map of the NSB of the region around the big city of Madrid (Spain). It is interesting to compare it with the previous works already mentioned in Section \ref{intro} around Perth, Dublin and Hong Kong. 

The area surveyed around Madrid is wider than any night sky brightness monitoring campaign with radial extensions up to 145 km from the centre of Madrid. Our data set covers an area of 5389 $km^2$ (assuming that the data points represent the night sky values of square cells of 2.2 $km^2$ side). The final map (when combining NSB taken with SQM photometers and radiance from satellite data) covers 268 $\times$ 210 km$^2$ (i.e. 56280 km$^2$). The number of useful data points is around 30,000. These numbers should be compared with (a) the 1957 night sky measurements in 199 locations and 1100 km$^2$ for Hong Kong monitoring campaign \cite{pun2012night}, (b) the 310 useful data points and around 40 km range of the study at Perth \citep{biggs2012measuring} with two radial tracks of 140 km and (c) the several hundreds of observations reported by \cite{espey2014initial}.

When comparing our survey around Madrid with the study of the Dublin area some coincidences are found in spite of the difference in size of the two cities. \citet[Figure 3]{espey2014initial} found a factor of $\sim 18\times$ in reduction of the night sky brightness for the first 60 km from Dublin that compares with the $\sim 16\times$ that we obtain in Madrid (from 18 to 21 $\;mag\;arcsec^{-2}$, see Fig.~\ref{fig:Radial_2}). 

The observation sites of the Hong Kong survey were located preferently in urban areas with only 14\% of the locations on rural settlements. \citet[][Figure 5]{pun2012night} shows an histogram of the NSB measurements collected with a mode around 16 $mag\;arcsec^{-2}$ and values of 20.1 $mag\;arcsec^{-2}$ for the darkest places. The distribution of NSB values for our study reaches its maximum at around 20.5-21.0 $mag\;arcsec^{-2}$ indicating that our surveyed areas had a significantly darker  sky (see Fig.~\ref{fig:Cells_Hist}).  

In the study around Perth, \cite{biggs2012measuring} found that the radial dependence of the NSB with distance could be fitted with an exponential of -1.93$\pm$0.07 for the long distance track (140 km) to the East. Perth is an isolated city and thus the result is similar to that predicted by the models \citep{walker1977effects,garstang1986model}, although the slope is less step than the -2.5 exponent of the Walker Law. However, Madrid and its surrounding region are full of small cities covering an extended area. This scenario is far from the model of a single population in the middle of an isolated area. This is why we found a slower exponent of -1.0 for distances up to 60 km and another one with a lower slope of -0.4 for greater distances.

\cite{biggs2012measuring} used interpolation (kriging) in order to create a map of NSB above Perth. This is a good choice when the data are sampled at random locations. In our study we have preferred to built the NSB map using a mesh of cells with the same size and position as that of the night-time {\it DMSP/OLS} imagery. Since the images of radiance ({\it OLS Stable Lights products}) are representative of clear and moonless nights \citep{elvidge1997mapping} the comparison between NSB and satellite nocturnal radiance data is straightforward. 

\section[]{Conclusions}

The spatial variation of the night sky brightness in a wide area around Madrid has been studied through observations made with SQM photometers on top of moving vehicles in order to facilitate the data gathering. The measurements were carefully filtered to avoid wrong observations caused by stray light and other unwanted effects. We have shown the feasibility of the method, which provides data to build maps of the night sky brightness at zenith. 

The map can be considered as representative of the average sky brightness in clear and moonless nights during the survey period (2010-2015). In foresight of future works it would be advisable to gather the data in a shorter span of time so as to obtain a snapshot of the situation. However the observations were instead performed during clear and moonless nights over a larger period to avoid the contamination caused by illuminated clouds and moonlight. Similar maps built with observations to be performed in the future will allow us to study the evolution of the NSB in this area.

The current map provides the spatial distribution of the night sky brightness at zenith at the centre of the Iberian peninsula around Madrid. It covers the Madrid metropolitan area and also regions far enough from the main polluting  sources. The study of the radial variation of the night sky brightness shows the expected effect of the darkening of the sky when moving from Madrid towards rural areas. We have found that this scenario is far from the simple model of a single bright and small city since Madrid is surrounded by a large and asymmetrical metropolitan area.   

The NSB map was built using a subset of clean data and has been binned in cells of 30 and 60 arcsec wide. This is the first time to our knowledge that a NSB map is presented in cells that corresponds to the mesh of the satellite imagery. This procedure make straightforward the comparison with radiance data, specially considering that the night sky brightness is closely related with the light pollution detected from space. Using this data we have been able to find a tight relationship between the night sky brightness map to the nocturnal radiance measured from space by the {\it DMSP} satellite ({\it OLS/DNB}). Using this fit, a prediction of the night sky brightness for the non-surveyed areas has been calculated in order to build a complete map without gaps. 

The results provide the essential ingredient to test the light pollution models that predicts night sky brightness using the location and brightness of the sources of light pollution and the scattering of light in the atmosphere, and also to gain insights into the nature of the diffuse light \citep{sanchezdemiguel2015diffuse}. The data presented in this paper is being used as reference calibration to build the new 'World Atlas of the artificial night sky brightness' \citep{falchi2015map}.

\section*{Acknowledgements}
This work has been partially funded by the Spanish MICINN (AYA2012-30717, AYA2012-31277 and AY2013-46724-P), by the Spanish program of International Campus of Excellence Moncloa (CEI), the Madrid Regional Government through the SpaceTec
Project S2013/ICE-2822, and by a FPU grant {\sl Formaci\'on de Profesorado Universitario} from the Spanish Ministry of Science and Innovation (MICINN) to Alejandro S\'anchez de Miguel. The support of the Spanish Network for Light Pollution Studies (Ministerio de Econom\'ia y Competitividad (MINECO) {\sl Acci\'on Complementaria} AYA2011-15808-E) is also acknowledged. The researchers have also used their own personal resources on this research and have been helped by relatives for some of the tracks:  Derlinda D\'iez, Julio Fern\'andez and Margarita Zamorano. We thank the volunteers that have contributed  with manual data acquisition: Marian L\'opez Cayuela, Roque Ruiz Carmona, Pablo Cepero, Daniel Escudero, Diego Pajuelo, Sara Bertr\'an de Lis, F\'elix Pradera, Javier S\'anchez, Sergio P\'erez Montalvo, David Cuesta, Pilar Pascual, Francisco Baca, Ada Garofano, Miguel \'Angel Carrete, Pilar Alcaraz,Almudena Jordana, Mercedes Turrero, V\'ictor Moreto, Beatriz Garc\'ia S\'anchez, Juan Carlos Larios, Ana Isabel Gonz\'alez, Roc\'io Luque, Ignacio C\'ardenas, Pedro Saura, Cristina Catal\'an and Jaime Izquierdo.

\section*{References}

\bibliographystyle{elsarticle-harv} 
\bibliography{references.bib}


\end{document}